\documentclass[
    ,final            
  ]
  {aipproc}

\layoutstyle{6x9}

\begin{document}

\newcommand{\chandra}{{\it Chandra}}
\newcommand{\spitzer}{{\it Spitzer}}
\newcommand{\snr}{G54.1+0.3}
\newcommand{\pwn}{3C~58}
\newcommand{\psr}{{PSR~J0205+6449}}

\def\ga{\ifmmode\stackrel{>}{_{\sim}}\else$\stackrel{>}{_{\sim}}$\fi}
\def\la{\ifmmode\stackrel{<}{_{\sim}}\else$\stackrel{<}{_{\sim}}$\fi}
\def\farcm{\hbox{$.\mkern-4mu^\prime$}}
\def\farcs{\hbox{$.\!\!^{\prime\prime}$}}

\title{Recent Progress in Studies\\ of Pulsar Wind Nebulae}


\author{Patrick Slane}{
  address={Harvard-Smithsonian Center for Astrophysics}
}

\begin{abstract}

The synchrotron-emitting nebulae formed by energetic winds from young
pulsars provide information on a wide range phenomena that contribute
to their structure. High resolution X-ray observations reveal jets and
toroidal structures in many systems, along with knot-like structures
whose emission is observed to be time-variable. Large-scale filaments
seen in optical and radio images mark instability regions where the
expanding nebulae interact with the surrounding ejecta, and spectral
studies reveal the presence of these ejecta in the form of thermal X-ray
emission. Infrared studies probe the frequency region where evolutionary
and magnetic field effects conspire to change the broadband synchrotron
spectrum dramatically, and studies of the innermost regions of the nebulae
provide constraints on the spectra of particles entering the nebula. At
the highest energies, TeV gamma-ray observations provide a probe of the
spectral region that, for low magnetic fields, corresponds to particles
with energies just below the X-ray-emitting regime.

Here I summarize the structure of pulsar wind nebulae and review several
new observations that have helped drive a recent resurgence in theoretical
modeling of these systems.

\end{abstract}

\maketitle


\section{Introduction}

It has long been known that the Crab Nebula is produced by the wind from
a young, energetic pulsar whose spin-down power manifests itself as a
synchrotron-emitting bubble of energetic particles.  Optical observations
reveal toroidal wisps and jet-like outflows near the pulsar. Faint
portions of a ring are observed, marking the region where the pulsar wind
undergoes a shock upon entering the surrounding nebula, the outer portions
of which are rich in filamentary structure from swept-up ejecta. Though
clearly formed in a supernova (SN) event in 1054~CE, the Crab shows no
trace of an associated supernova remnant (SNR), presumably indicating
that the SN blast wave is propagating through a very low density medium;
the nebula results entirely from the pulsar input (and its confinement
by surrounding material -- in this case, stellar ejecta) and is the
prototypical example of a pulsar wind nebula (PWN).

The diagram shown in Figure~1 illustrates the main points of the
most basic picture for a PWN: in an inner zone the wind flows away
from the neutron star (NS) with Lorentz factor $\gamma \sim 10^6$;
at a distance $R_w$ from the NS the wind passes through a termination
shock, decelerating the flow while boosting particle energies by
another factor of $\ga 10^3$; and beyond $R_w$ energetic electrons
in the wind radiate synchrotron emission in the wound-up toroidal
magnetic field to form the PWN which is confined at a radius $R_{PWN}$
by the inertia of the SN ejecta or the pressure of the interior of
a surrounding SNR. A detailed theoretical framework, incorporating
particle injection and diffusion, magnetic field evolution and
radiative and adiabatic losses, has been constructed within this
picture, allowing us to successfully predict and explain some basic
PWN properties (Reynolds \& Chevalier 1984; Kennel \& Coroniti
1984).  For a more detailed review on the structure and evolution
of PWNe, see Gaensler \& Slane (2006).

\begin{figure}[t]
  \includegraphics[width=\textwidth]{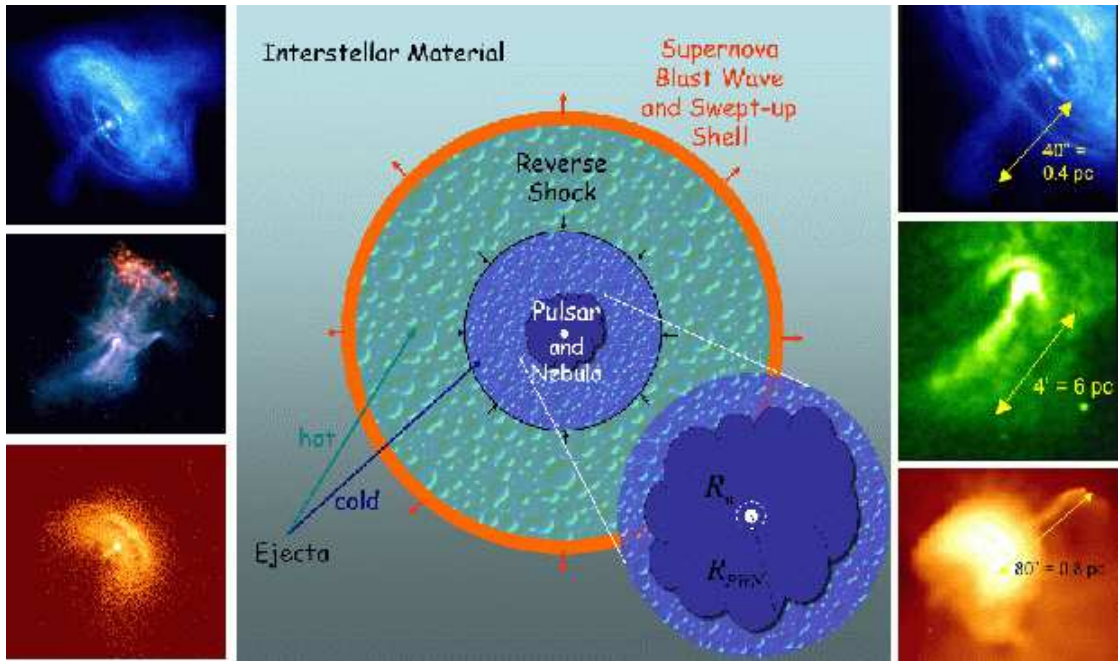}
  \caption{
Schematic diagram of a PWN within an SNR (see text for description).
\chandra\ images of the Crab Nebula, PSR~B1509$-$58, and Vela are shown
in the left panels. On the right are corresponding images of the jets in
these nebula, with indications of the relative sizes. Here and throughout,
north is up and east is to the left.
}
\end{figure}

Most models of PWNe have been built and tailored to explain the
properties of the Crab, but questions have lingered as to whether
or not other PWNe were really well-described by this ``prototype.''
Recent observations, particularly with high resolution X-ray
telescopes, have now shown that despite considerable differences
in age, environmental conditions, and pulsar properties, a large
number of PWNe share many of the Crab's traits.  In particular, it
has become startlingly apparent that pulsars release their energy
not in an isotropic fashion, but rather in equatorial winds and
polar outflows defined by the spin axis of the system. In Figure 1
(left panels) we show \chandra\ images of the Crab Nebula (Weisskopf
et al.  2000), PSR~B1509$-$58 (Gaensler et al. 2002), and the inner
nebula around the Vela pulsar (Helfand et al. 2001). Each shows a
well-defined axis of symmetry associated with the spin axis of the
pulsar, along with arc-like structures in the equatorial plane.
Similar structures have now been identified around a large number
of young pulsars.  The symmetry extends to large scales as well,
with enhanced confinement by the toroidal field producing nebulae
whose extent along the spin axis is larger than at the equator (see
Figure 2).  In the extended nebulae, particles suffer synchrotron
losses which gradually steepen the spectra with radius, as the
higher energy particles are burned off as they diffuse toward the
outer boundary -- an effect readily observed in X-ray spectra of
PWNe (e.g. Slane et al. 2001, Slane et al.  2004) that directly
probes the efficiency of the synchrotron cooling (Chevalier 2000).

\section{Jet/Torus Structure in PWNe}

As indicated above, when the free-flowing equatorial wind from the
pulsar encounters the more slowly-expanding nebula, a termination shock
is formed at the radius, $R_w$, at which the ram pressure of the wind
is balanced by the internal pressure of the PWN:
\begin{equation}
R_w = \sqrt{\dot E/(4 \pi \omega c \mathcal{P}_{\rm PWN})}, 
\label{eqn_ts}
\end{equation}
where $\omega$ is the equivalent filling factor for an isotropic
wind, and $\mathcal{P}_{\rm PWN}$ is the total pressure in the
shocked nebular interior.  Upstream of the termination shock, the
particles do not radiate, but flow relativistically along with the
frozen-in magnetic field. At the shock, particles are thermalized
and re-accelerated, forming a toroidal emission region in the
downstream flow. Emission from the termination shock region and the
downstream torus is seen clearly in the Crab Nebula (see upper left
in Figure 1); the geometry implied by the X-ray morphology is a
tilted torus, with a jet of material that flows along the toroid
axis, extending nearly 0.4~pc from the pulsar (see upper right in
Figure 1).  A faint counter-jet accompanies the structure, and the
X-ray emission is significantly enhanced along one edge of the
torus. Both effects are presumably the result of Doppler beaming
of the outflowing material, whereby the X-ray intensity varies with
viewing angle. Estimates of the nebular pressure are in good agreement
with the observed termination shock position based on the measured
$\dot E$ (see Eq. 1).

Similar geometric structures have now been observed in a number of
PWNe. \chandra\ observations of G54.1+0.3 (Figure 3), for example,
reveal a central point-like source surrounded by an X-ray ring whose
geometry suggests an inclination angle of about $45^\circ$ (Lu et al. 
2002).  The X-ray emission is brightest along
the eastern limb of the ring. If interpreted as the result of
Doppler boosting, this implies a post-shock flow velocity of $\sim
0.4c$. The ring is accompanied by faint bipolar elongations aligned
with the projected axis of the ring, consistent with the notion
that these are jets along the pulsar rotation axis.

The formation of these jet/torus structures can be understood as
follows (see Lyubarsky 2002).  Outside the pulsar magnetosphere,
the particle flow is radial. The rotation of the pulsar forms an
expanding toroidal magnetic field for which the Poynting flux
decreases with increasing latitude. Conservation of energy flux
along flow lines thus results in a variation of the wind magnetization
parameter, $\sigma$ (the ratio of Poynting flux energy density to
that in particles), and this anisotropy results in the toroidal
structure of the downstream wind.  Modeling of the flow conditions
across the shock shows that magnetic collimation produces jet-like
flows along the rotation axis (Komissarov \& Lyubarsky 2004).
This collimation
is highly dependent on the magnetization of the wind. At high latitudes
$\sigma$ is large, resulting in a small termination shock
radius and strong collimation, while close to the equator $\sigma$ is
much smaller and the termination shock extends to larger radii.

The jets observed in PWNe differ considerably from source to source,
spanning a large range in size (see right panels in Figure 1) and
in luminosity (relative to both that of the nebula and to the
available spin-down power of the pulsar).  At present, the connection
between the observed structure and the properties of the pulsar and
its environment is not well understood. The jets shown in Figure 1
all exhibit some amount of curvature, and this is typical of most
pulsar jets (e.g., that in 3C~58 -- see Figure 4). The hoop stresses
thought to confine the jets are subject to kink instabilities, which
can result in deflections of the flow, possibly explaining the
observed structures. The morphology of the Vela pulsar jet, shown
in Figure 1, is observed to change rapidly with time, consistent
with the presence of such instabilities.

\begin{figure}[t]
  \includegraphics[width=0.9\textwidth]{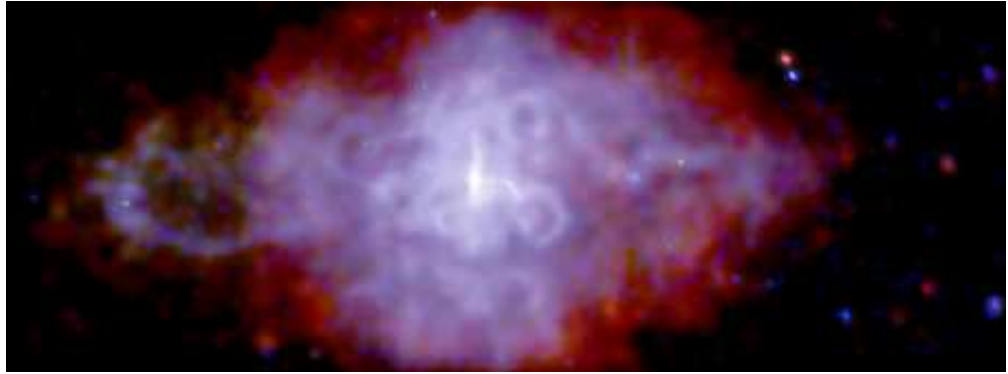}
  \caption{
\chandra\ image of 3C~58, from co-added images in the 0.5-1.0~keV
(red), 1.0-1.5~keV (green), and 1.5-10 keV (blue) bands.  The PWN
is elongated in the E-W direction, along spin axis defined by the
jet direction.  A shell of soft (red) thermal emission is evident,
as is a complex of synchrotron loops and filaments.
  }
\end{figure}

\begin{figure}[b]
  \includegraphics[height=.28\textheight]{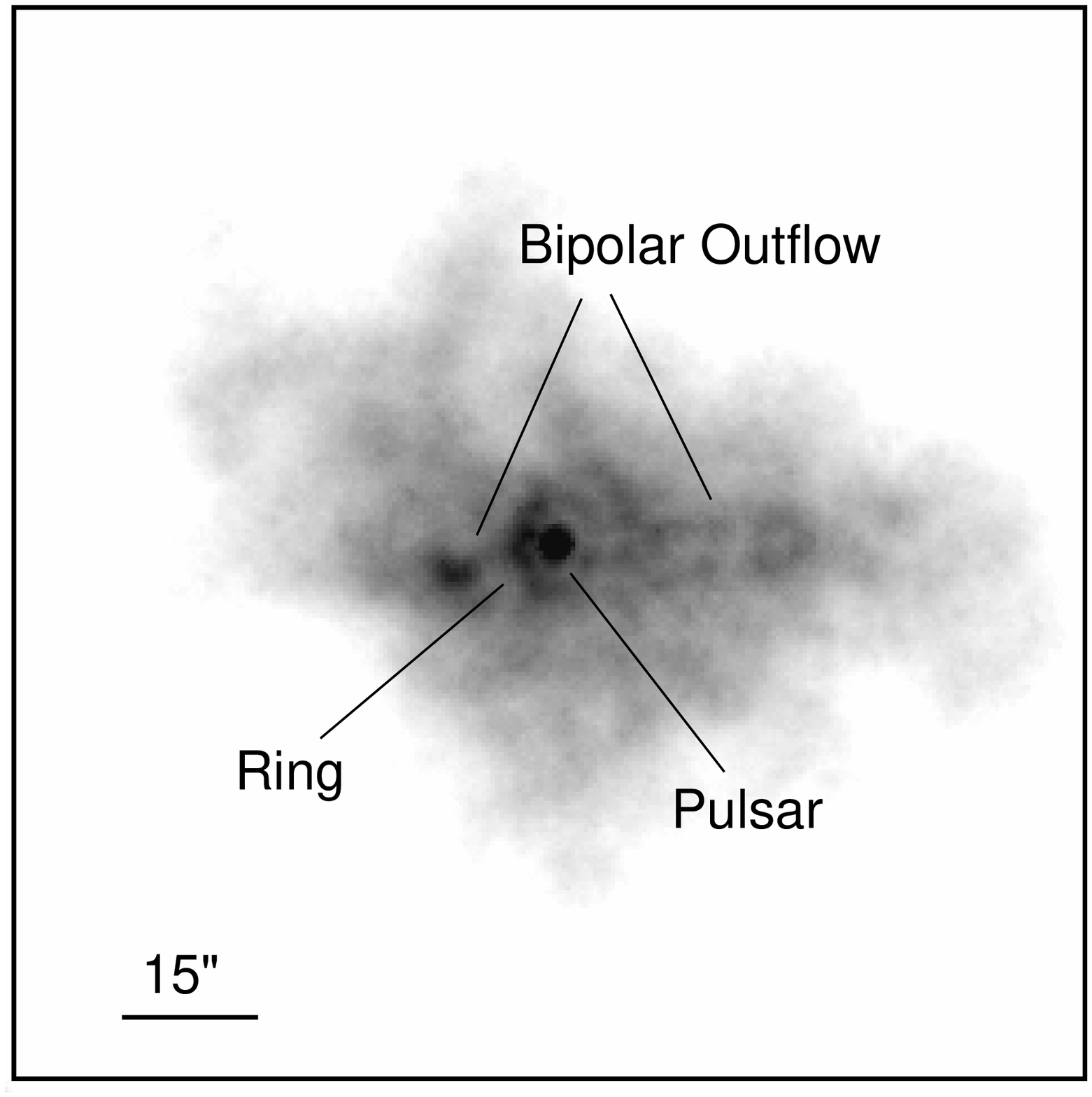}
  \includegraphics[height=.28\textheight]{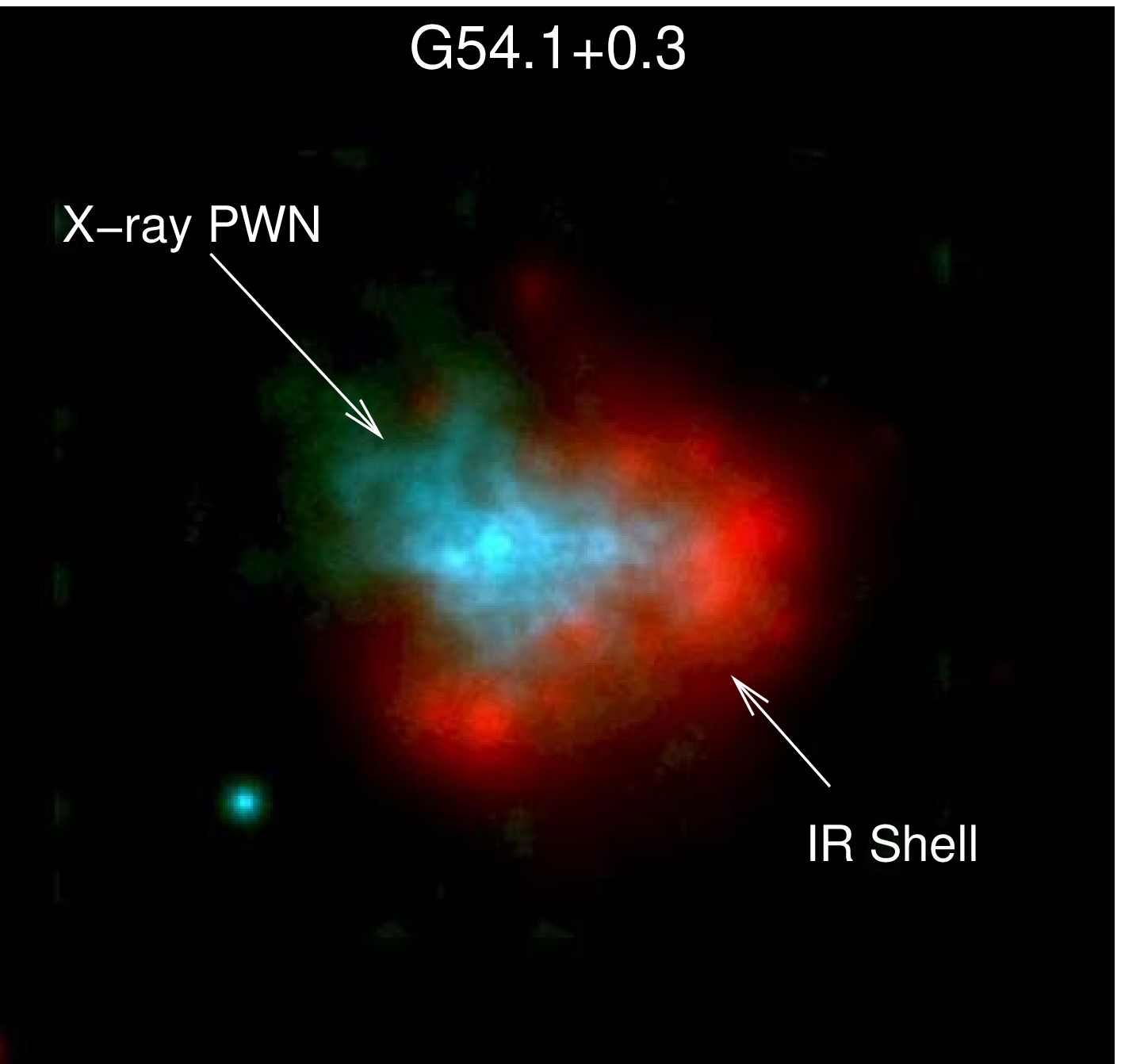}
  \caption{Left: \chandra\ image of G54.1+0.3 revealing a ring-like structure
surrounding the pulsar as well as jet-like outflows extending into a diffuse
nebula. Right: Composite showing \chandra\ image (blue) and \spitzer\
image (MIPS 24~$\mu$m). The IR observations reveal a shell in which the
PWN is embedded.
}
\end{figure}

\section{PWN Environments} 
While the SNR shell for the Crab Nebula has yet to be identified,
there is ample evidence for ejecta from its progenitor in the form of
Rayleigh-Taylor filaments formed as the relativistic gas in the nebula
expands into the slower-moving, denser ejecta.  These are clearly
seen in optical and infrared line maps of the Crab, and also in radio
observations where the synchrotron emission is enhanced along the
filaments due to compression of the magnetic field. There appears to
be little dust in the Crab's environment, however; \spitzer\ 
observations indicate a total dust mass of $< 1$~M$_\odot$, with
virtually no emission from small grains (Temim et al. 2006).
Searches for an
SNR shell continue, for the Crab and other young PWNe, because of the
considerable information such a detection would provide, in particular
on the composition and density profile of the ejecta through which the
young nebulae are expanding.

Optical studies of 3C~58 reveal filamentary structures associated
with ejecta (van den Bergh 1978; Rudie \& Fesen 2007), and X-ray
observations reveal a shell of shock-heated material with enhanced
abundances, suggesting an ejecta origin (Bocchino et al. 2001, Slane
et al. 2004, Gotthelf et al.  2006). This is illustrated in Figure
2 where we show a \chandra\ image that reveals extended low energy
emission enriched in Ne (shown in red) in the outer regions of the
PWN. The total mass of this material, along with the size of the
nebula, suggest an age of several thousand years (Chevalier 2005),
calling into question the often-assumed historical association of
3C~58 with SN 1181.

In addition to the soft X-ray shell, the \chandra\ observations of 3C~58
reveal a network of synchrotron loops and filaments spread throughout
the nebula whose nature is currently unknown. Unlike the radio filaments
seen in the Crab, these structures do not seem to correspond well with
the optical filaments in 3C~58. Kink instabilities can lead to loops
of magnetic flux being torn from the pulsar jet regions, and these
could manifest themselves as regions of enhanced synchrotron emission,
but more studies are required to determine whether these structures are
unique to 3C~58.

The outer regions of the diffuse X-ray emission in G54.1+0.3 show a
softer spectrum than the interior regions, but there is no evidence
for a thermal X-ray component associated with shock-heated ejecta or
ISM. {\it IRAS} observations reveal excess infrared emission above the
extrapolated synchrotron spectrum, however.  Recent \spitzer\ observations
(Slane et al., in preparation) reveal a shell of emission that
may be associated with the swept-up ejecta or shocked dust (Figure 3,
right). Of particular interest is a complex of infrared knots in the
western portion of the shell, directly in line with the termination
of the jet-like outflow seen in X-rays (Figure 3, left). Further IR
spectroscopy is required to determine the nature of the shell and its
association with G54.1+0.3.

\section{Broadband Spectra of PWNe}
The broadband spectrum of a pulsar wind nebula (PWN) is determined by
the input spectrum from the pulsar, as modified at the termination shock
where the pulsar wind joins the flow in the nebula, incorporating the
evolution of the nebula magnetic field strength as the nebula expands
into the ambient medium. For a power-law injection of particles
from the pulsar, a constant magnetic field in the nebula yields a power
law synchrotron spectrum with a break at a frequency
\begin{equation}
\nu_b \approx 10^{21} B_{\mu \rm G}^{-3} t_3^{-2}{\rm\ Hz}
\end{equation}
(where $B_{\mu \rm G}$ is the magnetic field strength, in $\mu$G, and
$t_3$ is the age in units of $10^3$~yr)
above which the synchrotron cooling time of the radiating particles is
less than the age of the nebula.
Additional contributions to the spectral structure in PWNe include the
effects of any time-dependent input from the pulsar, of the interaction
of SNR reverse shock with the PWN, and of any features in the injection
spectrum itself. 

\begin{figure}
  \includegraphics[width=\textwidth]{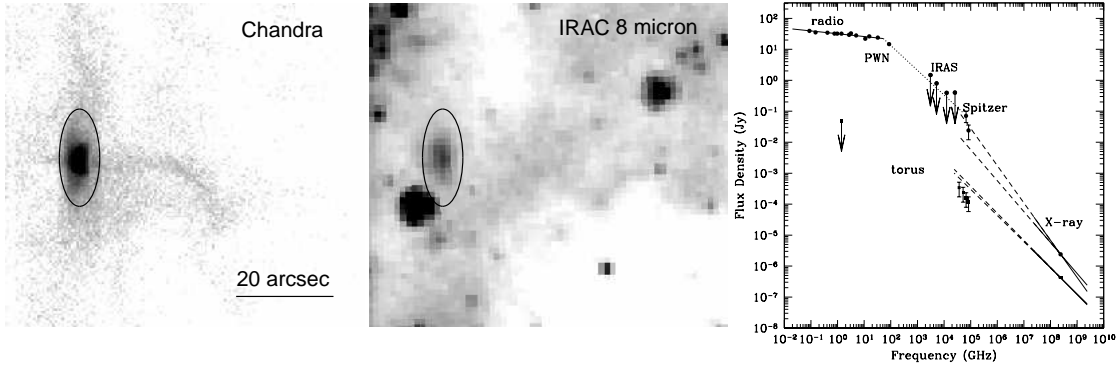}
  \caption{Left: \chandra\ image of the pulsar, torus, and jet in 3C~58.
Center: \spitzer\ image (IRAC 8~$\mu$m) revealing the 3C~58 torus.
Right: Broadband spectrum for 3C~58 and its torus, indicating that a broken
spectrum is injected into the nebula from the termination shock region.}
\end{figure}

Typically, PWN spectra are characterized by a flat power law index
at radio frequencies ($\alpha \approx -0.3$, where $S_\nu \propto
\nu^\alpha$) and a considerably steeper index in X-rays ($\alpha
\sim -1$). The Crab Nebula has a spectral break at $\sim 10^{13}$~Hz,
interpreted as a cooling break from a $\sim 300 \mu$G field, as well
as breaks at $\sim 3 \times 10^{14}$~Hz and $\sim 40$~keV. There is a
class of PWNe with breaks at much lower frequencies, however; 3C~58,
for example, has a break at $\sim 50$~GHz which, if interpreted as a
cooling break would imply a magnetic field well in excess of $1$~mG --
an unreasonably high field for a PWN, and one inconsistent with the fact
that the X-ray emission from 3C~58 extends all the way to the radio
edge of the nebula, indicating that no synchrotron loss breaks occur
far below the X-ray band.

\spitzer\ observations of 3C~58 (Slane et al. - in preparation)
show that the mid-IR emission from this PWN is consistent with
extrapolation of the X-ray spectrum. Of particular interest, though,
is that that pulsar torus is also detected in \spitzer\ observations
(Figure 4). The resulting spectrum requires at least one spectral
break, implying that the particle spectrum injected into the nebula
is not a single power law. Further characterization of the emission
in this region is of considerable importance for proper modeling
of the evolved nebular spectrum.

\section{High Energy Emission from PWNe}

Recent discoveries of very high energy $\gamma$-ray emission
associated with PWNe have opened a new channel for investigations
of the structure and evolution of these objects.  Of particular
interest is Vela~X, the large wind nebula associated with the Vela
pulsar.  It is located to the south of the pulsar, but this offset
position cannot be attributed to the pulsar velocity because the
direction of the pulsar motion, known from radio VLBI observations,
is not directed away from the center of the nebula.  Blondin et al.
(2001) explained Vela~X as a PWN that has been crushed and pushed
off-center by an asymmetric reverse shock wave that resulted from
the supernova interaction with an asymmetric surrounding medium.
{\it ROSAT} observations of the Vela SNR revealed a hard emission
component extending southward from the pulsar, into the Vela~X
region (Markwardt \& \"{O}gelman 1995) and a filamentary radio
structure extends alongside the same region.  HESS observations
reveal an extended region of TeV emission that follows this same
structure (Aharonian et al. 2006), presumably tracing filamentary
structure produced by the interaction of the PWN with the reverse
shock.

This emission in the TeV band probably originates from inverse-Compton
scattering of ambient soft photons with energetic electrons in the
nebula (Aharonian et al. 2006), although models have been considered
in which the emission is associated with the decay of neutral pions
produced in collisions of energetic ions with ambient hadronic
material (Horns et al. 2006).  If correct, the latter interpretation
would be of particular importance because it would demonstrate the
presence of ions in the pulsar wind.  Enhanced pion production from
such a wind may result from the mixing of ejecta material into the
PWN through instabilities produced by the reverse shock interaction.
Indeed, X-ray observations indicate the presence of shocked ejecta
material within the Vela~X region (LaMassa, Slane, \& de~Jager, in
preparation), strengthening the case for the reverse-shock interaction
scenario and potentially providing hadronic material with which an
ionic wind might interact to produce pions. However, further
observations of the high energy spectrum are required to differentiate
between the two emission scenarios.

Several recently-discovered HESS sources in the Galactic plane also
seem likely to be associated with PWNe. They are extended, have
spectra consistent with other PWNe, and have young radio pulsars
nearby. Examples include HESS J1804-216, HESS J1825-137,  and HESS
J1718-385 (Aharonian et al. 2005, Aharonian et al.  2007). For each
of these sources, the young pulsars suggested as the engines for
these nebulae are distinctly separated from the TeV centroids, possibly
suggesting that the PWNe have been disturbed by an asymmetric reverse
shock interaction as proposed for Vela X. In most cases, however, the
surrounding SNRs have yet to be identified.

In addition to the nature of the curious morphology for these
sources, a more vexing question centers on their very large sizes.
Several of these sources are observed to be extended on scales as
large as 15-30 arcmin or more. Using the dispersion-measure distances
of the pulsars which are suspected to have created the nebulae, the
physical sizes are in excess of 15 pc -- much larger than any known
PWNe observed in the radio or X-ray bands.  Diffusion of low energy
particles to distances far from the pulsar has been suggested as
an explanation for this large size, However, detailed modeling of
the environment that would lead to such a nebula, in the context
of the reverse shock interaction scenario, has yet to be carried
out. The results from these and other TeV observations are thus 
driving new investigations of PWN structure and evolution.

\section{Summary}
Recent advances in observations of PWNe have uncovered a wealth of
information regarding the axisymmetric nature of the wind and the
large- and small-scale structure of the nebulae. It is now clear
that the Crab Nebula is indeed the prototypical member of this class
-- albeit a particularly bright and energetic example; the elongated
nebula, swept-up ejecta, toroidal structure, and jets that characterize
the Crab are now being found in a multitude of PWNe. Observations
across the electromagnetic spectrum are helping to characterize the
evolution of PWNe, and opening new questions regarding the nature
of the wind and its interactions with its environs. These, in turn,
have stimulated new theoretical work on relativistic flows, shocks,
and jets with application to a broad range of topics in high energy
astrophysics.


\begin{theacknowledgments}
I would like to thank the conference organizers for the opportunity
to present this work, and also Bryan Gaensler, Jack Hughes, David
Helfand, Steve Reynolds, and my many other colleagues with whom I have
worked on studies of pulsar winds.
\end{theacknowledgments}

\end{document}